\documentclass{cpbtex}
\usepackage{siunitx}     
\usepackage{graphicx}    
\usepackage{amsmath}     

\usepackage[utf8]{inputenc}
\usepackage[T1]{fontenc}

\usepackage[final]{changes}

\usepackage{caption}

\begin{document}

\title{Micromagnet engineering for enhanced addressability of donor-cluster arrays in silicon}

\author{Hao Wang(王昊)$^{1,2}$\thanks{These authors contributed equally to this work.}, \  Bao-Long Zhao(赵宝龙)$^{1,2*}$, Shi-Hang Zhang(张世航)$^{1,2*}$, \\ Zi-Liang Jin(金子亮)$^{2,4}$, Huan Shu(舒焕)$^{2}$, Peihao Huang(黄培豪)$^{2}$,
\\ Guan-Yong Wang(王观勇)$^{2}$, Ming-Chao Duan(段明超)$^{2}$, Tian-Luo Pan(潘天洛)$^{3}$,
\\ Tao Xin(辛涛)$^{3}$, Guang-Chong Hu(胡光冲)$^{3}$\thanks{Corresponding author. E-mail: hugc@iqasz.cn}, Zhen Tian(田振)$^{3}$\thanks{Corresponding author. E-mail: tianzhen1@iqasz.cn}, \\ Da-Peng Yu(俞大鹏)$^{3}$, and Yu He(贺煜)$^{3}$\thanks{Corresponding author. E-mail: hey@iqasz.cn} \\
$^{1}${Southern University of Science and Technology, Shenzhen 518055, China.}\\  
$^{2}${International Quantum Academy, Shenzhen 518048, China.}\\ 
$^{3}${International Quantum Academy, and Shenzhen Branch,} \\ Hefei National Laboratory, Shenzhen 518048, China.\\
$^{4}${International Collaboration Laboratory of 2D Materials for Optoelectronics Science} \\  and Technology of Ministry of Education, Institute of Microscale Optoelectronics, \\ Shenzhen University, Shenzhen 518060, China.
}   


\date{\today}
\maketitle

\begin{abstract}
Phosphorus donors in silicon are one of the promising platforms for fault-tolerant quantum computing, owing to their long coherence times and compatibility with complementary metal-oxide-semiconductor (CMOS) technology. However, a key bottleneck for scaling this platform lies in the independent addressability of individual spin qubits. Here, we present a scalable quantum processor architecture that integrates scanning tunneling microscope (STM)-defined phosphorus donor-cluster arrays with on-chip micromagnets. By generating magnetic field gradients, the micromagnets enable a broader addressable frequency range compared to the addressing scheme based solely on variations in hyperfine (HF) interactions, which arise from the randomness of intra-cluster configurations. Through micromagnetic simulations to optimize micromagnet geometry, we demonstrate that these structures are compatible with customized array designs, facilitating the integration of over 120 atomic spin qubits and enabling frequency multiplexing with a single microwave antenna, which greatly reduces the overhead of \replaced{fan-out}{fanout} addressing lines. This design is further validated via magnetic measurements, which \added{generally} align with our micromagnetic simulations. These results establish a critical foundation for the development of scalable silicon quantum computing with atomic qubits.
\end{abstract}


\section{Introduction}
Phosphorus donors in silicon are a highly promising platform for fault-tolerant quantum computing, exhibiting exceptionally long relaxation times and coherence times,\cite{1,2,3,4,5} alongside high-fidelity single-qubit and two-qubit gate operations.\cite{6,7} Early efforts focused on single-atom donor configurations, as originally proposed by Kane.\cite{8} More recently, advancements in scanning tunneling microscope hydrogen depassivation lithography (STM-HDL) have enabled the creation of multi-donor quantum clusters,\cite{9, 10} culminating in the recent demonstration of an 11-qubit atomic processor in silicon.\cite{11, 12} This progress underscores the significant potential of donor-based qubits for scalability. However, as the number of qubits increases, the independent addressability of individual spins emerges as a critical bottleneck for realizing large-scale quantum computation.

In this work, we propose a quantum chip architecture that integrates STM-defined donor-cluster arrays with on-chip micromagnets, and systematically analyze the key factors influencing the addressability of spin qubits. We further simulate the magnetic field gradients generated by micromagnets and evaluate the associated enhancement of addressability. To validate this design, we fabricated cobalt (Co) micromagnet samples, measured their magnetic fields, and found the results were \replaced{generally}{closely} aligned with our simulations. 
\replaced{These results indicate that our proposal offers a potential solution to the frequency crowding issue}{These results confirm that our proposal mitigates the frequency crowding issue}, providing a practical pathway toward large-scale donor-based quantum computing.

\section{Addressability analysis and chip architecture}
The fundamental building block in the spin qubit chip is the donor cluster consisting of multiple phosphorus nuclei coupled to a single shared electron.\cite{10,11,12,13,14,15,16} The Hamiltonian of a single cluster can be expressed as $H = \gamma_e \bm B \cdot \bm S + \sum_i^{n} \gamma_n \bm B \cdot \bm{I}_i + \sum_i^{n} A_i \bm{S} \cdot \bm{I}_i$, where $\gamma_e = 27.97$ GHz/T and $\gamma_n = -17.23$ MHz/T are the gyromagnetic ratios of an electron spin and a nuclear spin, respectively; $\bm B$ is the external magnetic field; $\bm{S}$ and $\bm{I}_i$ are spin operators of the electron spin and the $i$-th nuclear spin, respectively; and $A_i$ is the hyperfine (HF) interaction strength between the electron spin and the $i$-th nuclear spin. When performing electron spin resonance (ESR) driving, the resonant frequency of the electron spin is conditioned on the nuclear spin state:\cite{13} 

\begin{equation}
f_{\mathrm{ESR}} = \gamma_e  B_0 + \sum_i^{n}(-1)^{m_i}\frac{A_i}{2},    
\label{eq:ESR}
\end{equation}
where the parameter $m_{i}=0$ (1) represents that the $i$-th nuclear spin is up (down) in the corresponding nuclear spin state, and $B_0$ is the static magnetic field. For the cluster-based system, the electron spin is initialized to the spin-down state. Thus, the driving frequency of nuclear magnetic resonance (NMR) for the $i$-th nuclear spin is $f_{\mathrm{NMR},i} = |\gamma_n B_0 - A_i/2|$. The single-qubit gate on the specific qubit is executed via NMR driving with the corresponding $f_{\mathrm{NMR},i}$. The multi-qubit controlled-$Z$ (CZ) gate (with multiple control qubits and one target qubit) can be implemented via the $\pi$ geometric phase accumulated during a 2$\pi$ electron spin rotation, conditioned on specific nuclear spin states.\cite{11,12,13,16} Additionally, Toffoli gates can be implemented by applying single-qubit Hadamard (H) gates in conjunction with multi-qubit CZ gates.\cite{12} Therefore, the electron-nuclear HF interaction plays a central role in the addressability of donor-based spin qubits, and its distribution mainly depends on the atomic configuration in the cluster, including inter-donor separation and crystallographic orientation.\cite{17} For example, small $A_i$ corresponds to a large inter-donor spacing, where the electron wavefunction becomes more delocalized, while larger $A_i$ indicates stronger electron wavefunction localization around the phosphorus nucleus. While STM-HDL enables atomic-level precision in device lithography, the spatial arrangement of phosphorus atoms within the clusters is inherently random following adsorption, incorporation, and overgrowth processes in silicon, which ultimately introduces considerable  HF interaction variations.\cite{13,18} The HF value of a 1-phosphorus (1P) in bulk is well-established at \SI{117.5}{MHz},\cite{19} and the measured values range from \SI{96.9}{MHz} to \SI{116.6}{MHz} for ion-implanted 1P donors, probably arising from the combined effect of variations in donor depths, electric fields, and strain.\cite{9,20} Multi-donor clusters exhibit a wide range of HF values (see Table 1), and the difference in HF coupling $\Delta A$ can provide sufficient addressability for a small number of qubits.\cite{10,11,13}

\begin{center}
{\footnotesize{\bf Table 1.} Experimental hyperfine energy values of P donors in STM-HDL-fabricated clusters.\\
\vspace{2mm}
\begin{tabular}{ccc}
\hline
Cluster type      & Hyperfine energy (MHz) & Total hyperfine energy (MHz)\\\hline
2P$^{\rm a)}$                & 83.0 $\sim$ 262.2 & 272.0 $\sim$ 524.4 \\
3P$^{\rm b)}$                & 6.0 $\sim$ 200.0 & 175.0 $\sim$ 318.0 \\
4P$^{\rm c)}$                & 1.0 $\sim$ 226.0 & 87.9 $\sim$ 465.3 \\
5P$^{\rm d)}$                & 0.6 $\sim$ 197.8 & 267.4 \\
\hline
\multicolumn{2}{l}{$^{\rm a)}$Ref.~\cite{9,21}; $^{\rm b)}$Ref.~\cite{10}; $^{\rm c)}$Ref.~\cite{11,12}; $^{\rm d)}$Ref.~\cite{11}.}
\end{tabular}}
\end{center}

As proposed in the scalable cluster-based spin qubit scheme,\cite{13} neighboring clusters are coupled to each other via inter-cluster exchange interaction, thus forming a cluster-array-based quantum processor with enhanced connectivity.
To quantify the HF-randomness-limited addressability of this multi-cluster system, we assume that all clusters are of the 3P type ($n=3$) and subsequently analyze the maximum number of distinguishable clusters under the impact of crosstalk errors. Crosstalk errors arise when ESR/NMR frequencies are insufficiently separated, and they can be estimated by\cite{21}

\begin{equation}
\label{eq:crosstalk}
\text{Error} = \frac{f_{\mathrm{Rabi}}^2}{f_{\mathrm{Rabi}}^2 + \Delta f_d^2},
\end{equation}
where $f_{\mathrm{Rabi}}$ is the Rabi frequency and $\Delta f_d$ is the driving frequency detuning between the resonant and the nearby transition frequencies. To underscore the impact of crosstalk errors, we take $f_{\mathrm{Rabi}} = \SI{1}{MHz}$, which is a sufficiently large value compared with experimental results.\cite{10,11,13} In this case, $\Delta f_{d} \geq \SI{10}{MHz}$ is necessary to achieve a crosstalk error rate below 1\%\added{, which can be further reduced by utilizing crosstalk suppression methods.\cite{12,13} We choose a larger $\Delta f_{d}$ to avoid overestimating the addressability due to the simplified estimation}. Based on the range of $A_{\text{3P}}$ derived from the tight-binding method,\cite{17} we adopt a HF coupling difference of $\Delta A = \SI{300}{MHz}$ for the following numerical evaluation. Ideally, the frequency band provided by the HF interaction is fully occupied by ESR frequencies separated by an interval of 10 MHz. In this scenario, the maximum number of distinguishable clusters ($N$) is estimated to be 10 according to

\begin{equation}
\label{eq:maxqubits}
N = \frac{\Delta A}{n \, \Delta f_d},
\end{equation}
which enables individually addressing a total of 30 qubits. The estimated maximum of \replaced{30 qubits}{10 distinguishable clusters} represents an upper bound for addressability solely provided via the randomness of HF. However, this number is readily achievable in practice for two reasons. Firstly, Eq. \eqref{eq:crosstalk} is derived for rectangular pulses; employing a Gaussian pulse allows crosstalk to be suppressed with a detuning $\Delta f_{d}$ substantially smaller than \SI{10}{MHz}. Secondly, the HF Stark coefficient can be varied from \SI{1.0}{MHz/(MV.m^{-1})} to \SI{11.2}{MHz/(MV.m^{-1})},~\cite{21} enabling considerable ESR frequency shifts via in-plane gates or top gates, as electric fields on the order of \SI{4.5}{MV/m} are readily attainable.~\cite{9} \added{Notably, the incorporation of exactly 3P is inherently probabilistic in practice. According to Kinetic Monte Carlo (KMC) simulations, the probability of forming a 3P cluster can be tuned to above 70\% by optimizing the STM lithographic window,  while clusters with fewer than two or more than three phosphorus atoms occur extremely rarely and can thus be neglected.~\cite{22} For 2P clusters, qubit addressability can still be achieved, given that the total hyperfine (HF) coupling is generally weaker than that of 3P clusters. Additionally, auxiliary clusters can be introduced into the array to enhance the robustness tolerance of the chip architecture against fluctuations in the number of dopant atoms, a point that will be revisited later.} Nevertheless, should any two qubits remain nearly degenerate despite the aforementioned methods---resulting in a crosstalk error exceeding 1\%---they can be safely idled by initializing both to the ground state and excluding them from the encoding space, and this mitigation strategy preserves the enhanced connectivity intrinsic to the donor-cluster array architecture.~\cite{12,13}


\begin{center}
\includegraphics{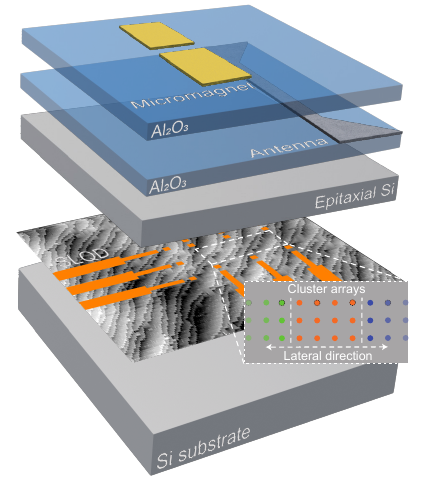}\\  
\captionof{figure}{\footnotesize Schematic illustration of the quantum processor architecture integrating micromagnets with phosphorus donor-cluster arrays patterned by STM-HDL. The planar nanostructures include phosphorus donor-cluster arrays, single-lead quantum dots, and other nanoscale circuit elements, which are embedded between a crystalline silicon substrate and an epitaxial silicon layer. \added{The clusters marked by black dashed lines are auxiliary clusters, which provide additional fabrication flexibility.} An Al$_2$O$_3$ insulating layer grown by ALD, a microwave antenna, a secondary Al$_2$O$_3$ insulating layer, and micromagnets are sequentially fabricated via \replaced{micro- and nanofabrication}{micro/nano-fabrication} processes.}
\label{fig:1}
\end{center}

When the number of clusters is further increased, the frequency crowding becomes a fundamental challenge for clusters with similar intracluster configurations. To overcome this problem, we propose introducing on-chip micromagnets that generate a magnetic field gradient across the donor-cluster array, which superimposes on the external magnetic field $\bm B_0$, shifting the resonant frequency of each electron or nucleus confined to the cluster according to its \deleted{lateral} position.
As illustrated in Fig.~\ref{fig:1}, the proposed architecture is built on a crystalline silicon substrate using STM-HDL technology. The planar structures encompass phosphorus donor-cluster arrays, single-lead quantum dots arrays serving as spin readout modules,\cite{23} and a versatile suite of gate arrays to precisely tune the chemical potential of the quantum dots, such as in-plane gates\cite{24} or epitaxial top-gates\cite{25,26}. These planar components are encapsulated with an epitaxial silicon layer, and ohmic contact electrodes are fabricated to establish electrical connections. Next, a dielectric layer (such as Al$_2$O$_3$, HfO$_2$) is deposited via atomic layer deposition (ALD), upon which an antenna is fabricated to enable qubit driving. Lastly, a second dielectric layer is grown to ensure electrical isolation, with micromagnets integrated above the predefined target array positions. \added{Since the in-plane gates are less than \SI{1}{\nano\meter} thick, the ohmic contact electrodes are more than \SI{5}{\micro\meter} away from the device core region, and the microwave antenna is designed to be laterally offset by hundreds of nanometers. Thus, the micromagnets (especially their corners) can be largely ensured to have no spatial overlap with these structures, thereby preventing the induction of surface irregularities in the micromagnets.} \added{Additionally, the introduction of a magnetic field gradient inherently renders qubits more susceptible to charge noise, leading to dephasing of the electron spin qubit. However, in the donor-based spin qubit scheme, the data qubits are nuclear spin qubits, while the electron spins are used only as ancilla qubits. Consequently, the impact of electron spin dephasing on the nuclear spin data qubits is indirect and mitigated. Moreover, except during the execution of ESR-based multi-qubit gates, the electron spins remain in the ground state during other quantum operations, further relaxing the constraints on electron spin coherence. Additionally, we can mitigate the negative impact of introducing magnetic field gradients on spin qubits by combining with dynamical decoupling techniques.\cite{27,28}}

The total HF coupling strength \(A_{\Sigma}\) of multi-donor clusters exhibits a broad distribution characteristic. For example, based on tight-binding numerical calculations, the \(A_{\Sigma}\) values of 3P donor clusters range from \SI{258}{MHz} to \SI{1050}{MHz}.\cite{29} We may use \(A_{\Sigma}=654\ \text{MHz}\) as the median for estimation, which corresponds to a magnetic field shift of $\Delta B_0 = A_{\Sigma}/ \gamma_e \approx \SI{23.4}{mT}$. As illustrated in Fig.~\ref{fig:1}, we consider arranging every ten clusters into a unit. Within each unit, the clusters are arranged in four columns with a ``3+2+3+2'' configuration. \added{Such a "3+2+3+2" configuration can be flexibly adjusted to address fluctuations in dopant number, such as "3+3+3+3" configuration.} These units are then tiled periodically. The typical cluster spacing is set to the commonly reported \SI{13}{nm},\cite{7,11} and the lateral extent of each unit is approximately \SI{52}{nm}. In the proposed design, a magnetic field gradient of approximately \SI{0.45}{mT/nm} is required to achieve a 654 MHz Larmor frequency shift between adjacent units. \added{Furthermore, this magnetic field gradient can shift the Larmor frequency between different columns within the same cluster group by approximately \SI{163.5}{\mega\hertz}, which also serves to enhance the addressability.}

\section{Micromagnet simulation and structural optimization}

\begin{center}
\includegraphics{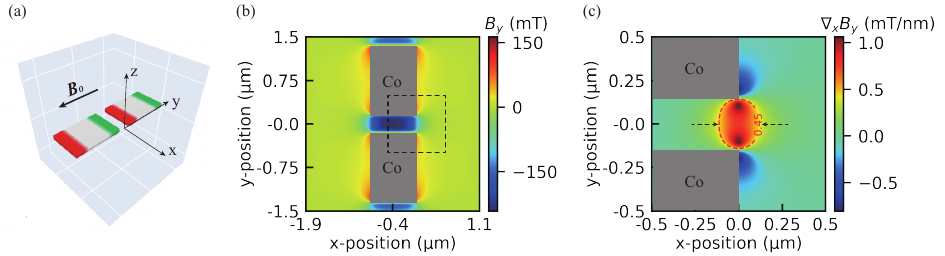}\\  
\captionof{figure}{\footnotesize Micromagnet geometry and magnetic field simulation. (a) Schematic illustration of the micromagnet, depicting the external magnetic field ($\bm B_0$), block-shaped magnetized micromagnets, and the Cartesian coordinate system. The coordinate origin is located at the midpoint of the gap and aligned along the extended edge of the micromagnets. (b) Simulated magnetic field y-component ($B_y$), where the dashed-box region represents the region of interest. (c) x-axis gradient of $ B_y$ ($\nabla_x B_y$) for the dashed-box region in panel (b). The red curve denotes a contour line with a value of \SI{0.45}{mT/nm} \added{and the range marked by the two black dashed arrows corresponds to the effective lateral range.} Gray cuboids in panels (b) and (c) denote the projected positions of the Co micromagnets.}
\label{fig:2}
\end{center}

To verify the feasibility of fabricating micromagnets suitable for the above-mentioned architecture, we performed magnetostatic simulations using the Python package Magpylib.\cite{30} A ``block'' geometry (consisting of two cuboids separated by a gap) was adopted in the simulation, with the coordinate origin positioned at the midpoint of the gap, as illustrated in Fig.~\ref{fig:2}(a). An external magnetic field ($\bm B_0$) was applied along the \added{negative} y-axis. 
The sampling plane was placed $\SI{40}{nm}$ below the micromagnets (corresponding to a stack of a $\SI{20}{nm}$ epitaxial silicon layer and a dielectric layer with a total thickness of $\SI{20}{nm}$). Ferromagnetic materials such as Fe, FeCo, and Co can be used for micromagnets~\cite{31}; among these, Co was chosen for its high saturation magnetization and widespread application in electron dipole spin resonance magnets.\cite{32} 
In the simulations, the saturation magnetization of Co was set to \SI{1.4}{MA/m}.\cite{31} The typical magnetic field of $B_y$ (Fig.~\ref{fig:2}(b)) and its gradients $\nabla_x B_y$ (Fig.~\ref{fig:2}(c)) can be acquired from the simulations. Figure \ref{fig:2}(c) delineates the region satisfying $\nabla_x B_y \ge$ \SI{0.45}{mT/nm} using a red dashed contour. \added{Since the gradient of $B_y$ along the y-direction is negligible in our proposed architecture (see Supplementary information for details), we define the extent occupied by this region along the x-axis as the effective range of the magnetic field gradient, which serves as the primary metric for evaluating the addressability of the micromagnet in our work.} Notably, significant magnetic field gradients are predominantly generated at the corners of the micromagnets.  

\begin{center}
\includegraphics{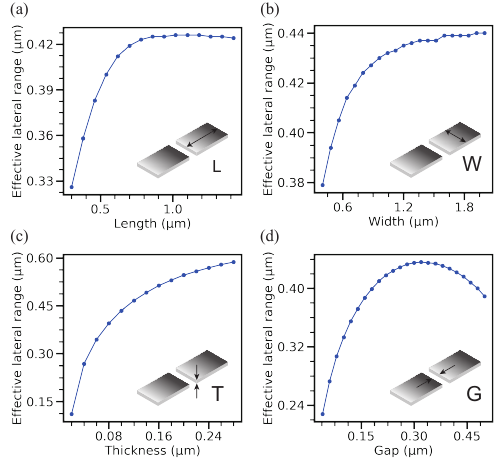}\\  
\captionof{figure}{\footnotesize Geometric optimization of micromagnets. The lateral \replaced{range}{extent} of the region with an effective magnetic field gradient of $\ge \SI{0.45}{mT/nm}$ is adopted as the optimization criterion. The dependence of this lateral \replaced{range}{extent} on the micromagnet length ($L$), width ($W$), thickness ($T$), and gap ($G$) is examined for panels (a), (b), (c), and (d), respectively. Only one geometric parameter is varied at a time, with all other parameters fixed to their default values: length $L = \SI{1.2}{\micro m}$, $W = \SI{0.8}{\micro m}$, $T = \SI{0.1}{\micro m}$ and $G = \SI{0.3}{\micro m}$.}
\label{fig:3}
\end{center}

To maximize the effective range of the magnetic field gradient, the geometric shape of the micromagnets was optimized by tuning one geometric parameter (length \(L\), width \(W\), thickness \(T\), or gap \(G\)) while keeping the remaining parameters fixed at their default settings: $L = \SI{1.2}{\micro\meter}$, $W = \SI{0.8}{\micro\meter}$, $T = \SI{0.1}{\micro\meter}$, and $G = \SI{0.3}{\micro\meter}$. As illustrated in Fig.~\ref{fig:3}, the effective magnetic field gradient range increases monotonically with $L$, $W$, and $T$, albeit with diminishing growth rates, which can be primarily attributed to the volume effect of the micromagnet. In contrast, the dependence on $G$ is non-monotonic: the range first increases and then decreases, reaching a maximum at $G = \SI{0.2}{\micro\meter}$.\cite{31} This is attributed to the superposition effect of the magnetic field gradients generated by the two corners of the micromagnet. When the gap is small, the ranges of the magnetic field gradients produced by the two corners exhibit significant overlap. As the gap increases, the overlapping region gradually decreases, while the total effective range progressively expands and begins to decline after reaching a critical point. Considering practical microfabrication yield constraints, the optimized geometric parameters were determined as $L = \SI{1.2}{\micro\meter}$, $W = \SI{0.8}{\micro\meter}$, $T = \SI{0.1}{\micro\meter}$, and $G = \SI{0.3}{\micro\meter}$. 
These parameters yield an effective gradient range of approximately \added{\SI{222}{nm}} \deleted{\SI{231}{nm}} along the x-axis, which is more than four times larger than the length of the aforementioned 10-cluster unit along the x-axis, \replaced{allowing}{thus enabling} the independent addressability of over 120 phosphorus atoms. 
Furthermore, the number of addressable spin qubits can be further enhanced by increasing the thickness of the micromagnets. 

The addressability of 120 atomic qubits is enabled by multiplexing a single ESR/NMR antenna. This approach greatly alleviates the routing overhead for individual addressing lines. The structure shown in Fig.~\ref{fig:1} functions as a fundamental module in a quantum processor. With hundreds or thousands of such modules coupled together via cavity quantum electrodynamics\cite{33} or qubit shuttling techniques\cite{34,35}, in turn, it would be feasible to realize a scalable, fault-tolerant quantum computer integrating millions of qubits.


\section{Sample preparation and magnetic characterization}

To validate the simulation results, we fabricated micromagnet samples for experimental testing. Sample preparation mainly involved micromagnet fabrication and magnetization. The micromagnets were fabricated using electron-beam lithography combined with a lift-off process. A double-layer resist system was employed to create the necessary undercut profile for a clean lift-off result, consisting of PMMA-A2 or ZEP as the bottom layer and PMMA-A5 as the top layer. The micromagnet patterns were exposed at an accelerating voltage of \SI{15}{\kilo\volt} using an optimized dose of 
\SI{14}{\micro C/cm^2} to ensure conformance with the design specifications. The exposed resist was developed in an $\text{MIBK:IPA} = 1\colon3$ for $60\ \text{s}$, followed by rinsing \replaced{in IPA}{inIPA} for $20\ \text{s}$ and drying under $\text{N}_2$ flow. A Co film with a thickness of approximately $100\ \text{nm}$ was deposited onto a silicon substrate using electron beam evaporation at a rate of \SI{0.3}{\nano\meter/\second}. The final structures were obtained after removing the remaining resist and excess metal in an acetone bath, as illustrated in Fig.~\ref{fig:4}(a). The Co micromagnet magnetization was performed on a Quantum Design PPMS DynaCool. The magnetic field was applied along the in-plane $y$-direction, ramped up/down at $10\ \text{mT/s}$, and held at \SI{8000}{\milli\tesla} for 5 minutes at room temperature to ensure that the Co micromagnet magnetization is fully saturated.\cite{36,37} \deleted{Alternatively, the magnetization process of qubit samples integrated with micromagnets can also be achieved in situ using the superconducting magnets available in dilution refrigerators.}

Stray magnetic fields generated by \deleted{the} Co micromagnets were imaged using a commercial scanning nitrogen-vacancy center microscope (SNVM, CIQTEK) via optically detected magnetic resonance (ODMR).\cite{38,39} \added{Since the detection range of SNVM (typically <10 mT) is much smaller than the average magnetic field strength at 40 nm from the micromagnet,} NV centers were located approximately $350\ \mathrm{nm}$ above the Co micromagnet to avoid fluorescence quenching induced by large stray fields.\cite{38} \added{
More simulation results regarding the magnetic field distribution on planes at different heights, as well as their evolution characteristics, are presented in the Supplementary Information. In future work, we plan to employ alternative characterization techniques, such as magnetic force microscopy (MFM) or scanning SQUID microscopy (SSM). These approaches would provide complementary insights into the magnetic domain structures and stray field distributions, allowing us to acquire more comprehensive information about the actual magnetic landscape at the nanoscale.} The full-range ODMR spectrum of the NV center was recorded during frequency sweeping to acquire the magnitude of the stray field along the NV center quantization axis ($\bm B_{\text{NV}}$). We obtained all three magnetic field components ($B_x$, $B_y$, and $B_z$) using the method reported in Ref.~\cite{40}, and these results \added{using bilinear interpolation}  are in general agreement with micromagnetic simulations, as shown in Fig.~\ref{fig:4}(b). \added{In accordance with the standard magnetostatic laws, we can indirectly infer that the magnetic field distribution at the qubit location is also generally consistent with the simulation results.} However, the experimental results exhibit a non-ideal local distribution and a smaller strength of the stray magnetic field, \replaced{which may be attributed to the following factors:\cite{30,41} First, the formation of a multi-domain structure (accompanied by spin relaxation) is likely to occur after the magnetization process, which causes a significant reduction in the remanent magnetization ($M_\text{r}$) relative to the saturation magnetization ($M_\text{s}$), as opposed to the idealized perfect single-domain model adopted in our simulations. Second, the evaporated cobalt film has a polycrystalline structure with randomly oriented magnetic easy-axis in individual grains; this intrinsic structural characteristic inherently gives rise to the observed local magnetic inhomogeneity. Third, minor factors including surface oxidation (which forms a non-magnetic dead layer), edge roughness, and lift-off residues also induce local magnetic field distortions and the formation of magnetic pinning sites. These inhomogeneity-inducing factors affect the addressability to a certain extent, since the strength of the magnetic field gradient at certain positions may fail to meet the design specifications.}{which may be attributed to the polycrystallinity and nonuniform magnetization of the Co micromagnet.\cite{27,38}} In the future, more efforts should be devoted to optimizing micromagnets, including the selection of magnetic materials, minimization of surface roughness, and magnetic domain structure simulations\cite{31,32,42}, as well as the achievement of their high-quality integration with donor-cluster array chips.\cite{13}

\begin{center}
\includegraphics{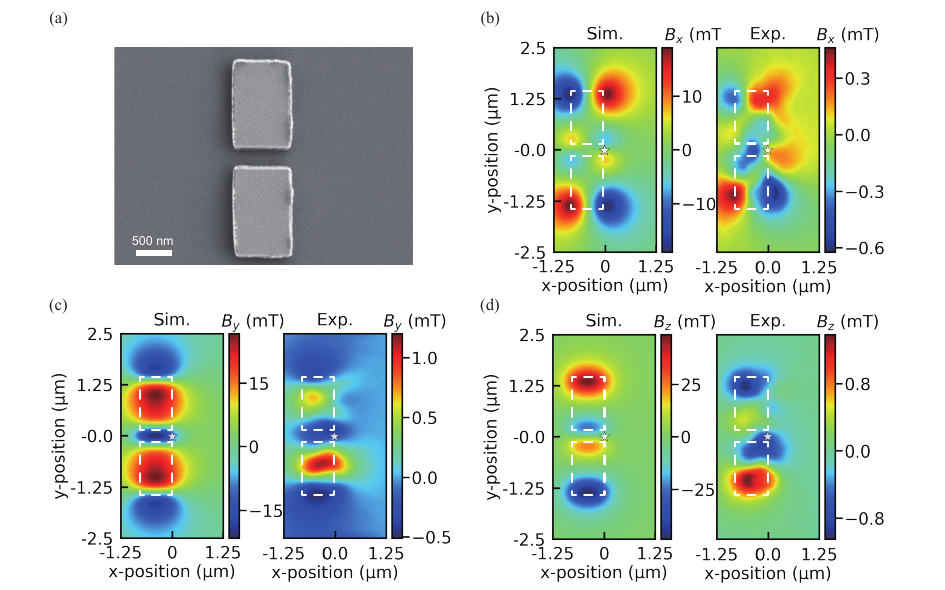}\\  
\captionof{figure}{\footnotesize Co micromagnet characterization. (a) Scanning electron microscopy (SEM) image of a \replaced{cobalt}{coblt} micromagnet with measured dimensions: $L = \SI{1.216 \pm 0.001}{\micro\meter}$, $W = \SI{0.811 \pm 0.013}{\micro\meter}$, $G = \SI{0.280 \pm 0.007}{\micro\meter}$, and a deposited thickness of $100\ \mathrm{nm}$. Comparison between simulated and SNVM-measured magnetic field components: $B_x$ in \textbf{(b)}, $B_y$ in \textbf{(c)}, and $B_z$ in \textbf{(d)}. \added{The experimental data is visualized using bilinear interpolation to represent the continuous magnetic field distribution.} The micromagnets are outlined by dashed lines, and the quantum dot arrays are marked with light gray pentagrams at their respective positions.}
\label{fig:4}
\end{center}

\section{Conclusion}
To address the critical addressability challenge of phosphorus donors in silicon for quantum computing, we propose integrating micromagnets with STM-fabricated donor-cluster arrays. Through systematic structural optimization of block-type Co micromagnets, simulations demonstrate that a lateral range of \SI{231}{\nano\meter} is attained on the qubit plane, within which the magnetic field gradient reaches a minimum of 0.45 mT/nm. By leveraging the variations in HF interactions of donor clusters and the gate-induced Stark effect, the proposed architecture enables the addressability of over 120 qubits for one modular structure with a single ESR/NMR antenna. Measurements of the micromagnets are in \replaced{general}{good} agreement with simulation results, further validating our design. This work establishes a critical foundation for advancing large-scale silicon-based quantum processors. Broader integration potential can be unlocked in the future through more sophisticated atomic engineering and micromagnet engineering, combined with long-distance qubit-coupling techniques.

\section*{Data availability statement}
The datasets generated and/or analyzed during the current study are available from the corresponding author upon reasonable request.

\addcontentsline{toc}{chapter}{Acknowledgment}
\section*{Acknowledgment}
This work was supported by the National Natural Science Foundation of China (Grants No. 92165210, 62174076, 11904157, 12275117, 92565304), the Innovation Program for Quantum Science and Technology (2021ZD0302300，2024ZD0300400), Guangdong Basic and Applied Basic Research Foundation (2022B1515020074), Guangdong Major Project of Basic Research (2025B0303000007), Shenzhen Science and Technology Program (Grants No. KQTD20200820113010023).

We acknowledge the Chen Ting-Yong group for their valuable assistance with the magnetizing process of micromagnets using the PPMS instrument.

\end{document}